\documentclass[reqno,aps,prd,twocolumn,showpacs,showkeys,preprintnumbers,amsmath,amssymb,amsfonts]{revtex4}
\newcommand{\wh}{\widehat}
\usepackage{graphics}
\usepackage[dvips]{graphicx}
\usepackage{dcolumn}
\usepackage{bm}
\usepackage[bookmarksnumbered]{hyperref}
\usepackage{color}

\numberwithin{equation}{section}

\begin{document}

\preprint{PA1.4a}

\pdfbookmark{Pioneer anomaly: a drift in the proper time of the spacecraft}{tit}
\title{Pioneer anomaly: a drift in the proper time of the spacecraft}


\author{Vikram H. Zaveri}
\email{cons\_eng1@yahoo.com}
\affiliation{B-4/6, Avanti Apt., Harbanslal Marg, Sion, Mumbai 400022 INDIA}



\date{May 13, 2008}

\begin{abstract}
A relativistic theory is proposed to explain the anomalous accelerations of Pioneer 10/11, Galileo and Ulysses spacecrafts. The theory points out at the limitations of the weak field approximation and proposes a drift in the proper time of the spacecraft outside the framework of general relativity. The theory yields a very accurate and precise value for the anomalous acceleration. In this theory the proper time of a body is associated with the gravitational frequency shift of the constituent fundamental particles of the body. The frequency shift changes the energy level of the body which gets reflected in its relativistic mass and therefore in its motion. This change in energy level causes the time like geodesics to deviate from that of the standard theoretical models. We introduce proper time in the line element of a metric theory according to a fixed set of rules laid down by general relativity for introducing deviation in the flat Minkowski metric. The frequency shift for bodies of different composition traversing different trajectories however, is not the same and this gets reflected in its motion as an unmodeled anomalous effect. This association of proper time with the gravitational frequency shift of the body requires the flat Minkowski metric to deviate in different ways for different two body systems. This solves the problem of anomalous acceleration in a very simple way. The solution to Pioneer anomaly given here yields anomalous acceleration  within the limits of observational accuracy. Gravitational redshift of light, bending of light and perihelic precession of planets are within the permissible limits. The theory shows that Einstein's field equations do provide some clue to the Pioneer anomaly but the solution is not very accurate. Hence the need to go beyond general relativity. The theory can also explain the rotation curves of the spiral galaxies.
\end{abstract}
\maketitle
\section{Introduction}
\hspace*{5 mm} Among many unexplained phenomeon of physics \cite{1}, the Pioneer anomaly is one of the foremost and quite well established. Very accurate navigation of the Pioneer 10/11 spacecraft was limited by a small, anomalous frequency drift of their carrier signals received by the radio-tracking stations. This drift between the observed signal and the modeled frequency using deep space navigational codes was interpreted as an approximate constant sunward acceleration and has come to be known as the Pioneer anomaly \cite{2,3,4,5,22,24,27,28}. Similar drifts were also observed during the navigation of Galileo and Ulysses spacecrafts \cite{3}. Pioneer 10/11 data supports the presence of this anomalous acceleration between the heliocentric distances of 20 AU and 70 AU. The value of this acceleration is $a_p=(8.74\pm1.33)\times10^{-10}m/s^2$. Galileo measurement showed $a_p=(8\pm3)\times10^{-10}m/s^2$ upto a distance of 5 AU. Ulysses data showed $a_p=(12\pm3)\times10^{-10}m/s^2$ over a heliocentric distance variation from 5.4 to 1.3 AU. These values were established after accounting for all known sources of systematic errors. Galileo and Ulysses navigation data are not considered very reliable for use in an independent test of the anomaly.\\
\hspace*{5 mm} At present the origin of Pioneer anomaly is unclear. The cause could very well be an overlooked systematic error on board the spacecraft. There is also a possibility of involvment of new physics. To address the later, several solutions have been proposed. Brownstein and Moffat \cite{6} have proposed a relativistic modified gravitational theory including a fifth force skew symmetric field. The theory allows for a variation with distance scales of the gravitational constant G. The theory has a repulsive Yukawa force added to the Newtonian acceleration law for equal "charges". Page, et al. \cite{7} have proposed that the minor planets provide an observational vehicle for investigating the gravitational field in the outer solar system, and that a sustained observation campaign against properly chosen minor planets could confirm or refute the existence of the Pioneer Effect. Iorio and Giudice \cite{8} have investigated the effects that an anomalous acceleration as that experienced by the Pioneer spacecraft would induce on the orbital motions of the Solar System planets beyond 20 AU threshold and ruled out the possibility of existence of an anomalous force field in the region 20-40 AU. Exirifard \cite{9} has proposed that a generally covariant correction is the cause of the observed Pioneers' anomaly. Scheffer \cite{10} points out that this unmodelled acceleration (and the less well known, but similar, unmodelled torque) can be accounted for by non-isotropic radiation of spacecraft heat. This paper considers new sources of radiation, all based on spacecraft construction and contends that the entire effect can be explained without the need for new physics. Ra$\tilde{n}$ada \cite{11} proposed that the observed anomalous effect is due to an adiabatic acceleration of the light and can be explained using the same phenomenological Newtonian model which accounts for the cosmological evolution of the fine structure constant. He has argued that the spaceships might not have any extra acceleration but would follow instead the unchanged Newton laws. In \cite{12,13,14} Ra$\tilde{n}$ada proposed that the anomalous phenomenon would be due to a cosmological acceleration of the proper time of bodies with respect to the coordinate time, and in \cite{30} he proposed the relative march of the atomic clock-time of the detectors with respect to the astronomical clock-time of the orbit as the cause of the observed anomaly. Rosales et al. \cite{15,16} also proposed the cosmological expansion model to explain the effect. A scalar potential model \cite{25} claims that matter causes a static, non-flowing warp of the $\rho$ field that causes the pioneer anomaly. In SPM, $\rho$ field exists between spiral galaxies and elliptical galaxies. Reynaud and Jaekel \cite{31} proposes metric extensions of general relativity to solve pioneer anomaly. The list is by no means exhaustive. The theory presented in following sections is based on earlier work called periodic relativity (PR) \cite{17} where I found it necessary to redefine the orbital energy of the body and eliminate weak-field approximation from the solution to Einstein's field equations.

\section{Physics behind the present theory}
\hspace*{5 mm} In PR \cite{17} it was proposed that the time is a periodic phenomenon and can only be defined as period between two events. The most fundamental aspect of time is the period of a wave. With this concept we were able to derive the gravitational redshift of light without utilizing Riemannian geometry and geodesic trajectories. The weak field approximation was not necessary and instead of proper time of light, only the frquency of light was used throughout the derivation. As a result PR does not compromise the constancy of velocity of light. In order to derive the bending of light in gravitational field we introduced tangential and normal (to the particle trajectory) components of the Newtonian radial acceleration in the Newtonian orbital energy equation. These components act along the line of directions of tangential and normal components of the velocity vector of the orbititng body. The normal component of the velocity vector however is absent. This introduced a factor $\left(\cos{\psi}+\sin{\psi}\right)$ in the formula for acceleration which yields geodesic like trajectories and correct value for the bending of light. Here $\psi$ is the angle between the radial vector and the tangential component of the velocity vector. This factor makes the orbital energy equation invariant but it does not alter the resultant radial acceleration which is same as Newton's theory.
\begin{align}\label{2.1a}
\sqrt{\left(\frac{\mu}{r^2}\cos{\psi}\right)^2+\left(\frac{\mu}{r^2}\sin{\psi}\right)^2}=\frac{\mu}{r^2},
\end{align}
We also proposed that the gravitational attraction exists between the relativistic masses and not just the rest masses. This when written in the form of a principle would mean that the gravitational mass is equal to the relativistic mass. Accordingly, in the coordinate system of a central mass, the gravitational attraction would exist between the rest mass of the central body (ignoring the spin) and the relativistic mass of the orbiting body. This proposal lead us to derive the higher order term for the orbital period derivative of a binary system \cite{18} which is very close to the currently accepted value of Blanchet and Sch$\ddot{a}$fer \cite{19}.\\
\hspace*{5 mm} We defined a general form of Max Planck's quantum hypothesis applicable to massive particles and derived a relation that showed that when a force acts on a fundamental massive particle, it accelerates the particle and simultaneously alters its associated de Broglie frequency and therefore force consists of two components, the Lorentz force and what we called the de Broglie force responsible for effecting the change in the associated frequency of massive particle \cite{17}. When de Broglie force is ignored, the invariant relationship between force and energy is destroyed. For photon trajectory in a gravitational field, we made use of this de Broglie force component to derive the gravitational redshift of light without making use of the weak field approximation and we made use of the Lorentz force component to derive the deflection of light, again without making use of the weak field approximation. In case of a gross body such as a planet, we can conviniently deal with the Lorentz force but it is not possible to directly account for the de Broglie force. To compensate for this deficiency, in PR we make use of the concept of proper time, but in general relativity the things are somewhat mixed up because of the weak field approximation.
In PR the ratio of coordinate time interval and proper time interval 
for photon is exactly one \cite{17}. This is not the case 
with general relativity. Therefore velocity of light in PR is perfectly 
constant like special relativity. But in general relativity it is not constant. 
In GR when photon travels through the gravitational field, its proper time interval always remains zero and  
does not vary in the same proportion as its period. In PR the 
proper time interval varies exactly in the same proportion as its period. 
And since this proportion is exactly 1 for the time interval corresponding to single
wavelength, the poper time interval can simply be replaced by the period. 
So this is the connection between the proper time interval and photon frequency. 
As the photon frequency shifts, the proper time interval drifts.
Hence PR proposes a definite connection between the proper time of a gross body and the associated de Broglie wavelengths and frequencies of its constituent fundamental particles which are in bound states. The new theory yields a line element which satisfies Einstein's field equations but is at variance with the weak field approximation and Schwarzschild solution. Introduction of these concepts in to Newtonian theory of gravity lead us to modify the commonly used expression relating the radial acceleration and the acceleration of the orbiting body which is   
\begin{align}\label{2.1b}
\frac{d^2\mathbf{r}}{dt^2}=\frac{d\mathbf{v}}{dt}.
\end{align}
The new equations we proposed are
\begin{align}\label{2.1c}
\frac{d\mathbf{v}}{dt}=\left[\frac{d^2s}{dt^2}\mathbf{\wh T}+\kappa\left(\frac{ds}{dt}\right)^2\mathbf{\wh N}\right]=
\frac{d^2\mathbf{r}}{dt^2}\left(\cos{\psi}+\sin{\psi}\right),
\end{align}
\begin{align}\label{2.1d}
\frac{d^2\mathbf{r}}{dt^2}=
-\frac{\mu}{r^2}\left(1-\frac{v^2}{c^2}\right)\mathbf{\hat r}\approx-\frac{\mu}{r^2}(\mathbf{\hat r}),
\end{align}
\begin{align}\label{2.1e}
\mathbf{v}\mathbf{F}=v\left[m\sqrt{\left(\frac{d^2s}{dt^2}\right)^2+\kappa^2\left(\frac{ds}{dt}\right)^4}+\frac{hv}{c^2}\frac{d\nu}{dt}\right].
\end{align}
It needs to be understood that $d^2\mathbf{r}/dt^2$ is a radial vector but $d\mathbf{r}/dt$ is not a radial vector which acts along the velocity vector $\mathbf{v}$. Therefore the conversion factor $\left(\cos{\psi}+\sin{\psi}\right)$ does not play any role in this expression of velocity $\mathbf{v}=d\mathbf{r}/dt$ which remains unaltered. The effect of factor $\left(\cos{\psi}+\sin{\psi}\right)$ would be significant for bodies in transition from near circular orbits to near radial motion. Much of this effect goes towards the formation of trajectory curvature. For near circular or near radial motions this factor is almost unity. The angle between the radial vector and the velocity vector $\psi$ is defined as 
\begin{align}\label{4.40c}
\psi=\tan^{-1}{\frac{r}{\dot{r}}} \qquad where \qquad \dot{r}=\frac{dr}{d\theta}.
\end{align}
If we introduce $u=1/r=(\mu/h^2)(1+e\cos{\theta})$ we get
\begin{align}\label{4.40ca}
\psi=\tan^{-1}\left[\frac{1+e\cos{\theta}}{e\sin{\theta}}\right]. 
\end{align}
\begin{align}\label{4.40d}
\sin{\psi}=(r/\dot{r})/\sqrt{(r/\dot{r})^2+1}.
\end{align}
\begin{align}\label{4.40e}
\sin^2{\psi}=(r/\dot{r})/\sqrt{(r/\dot{r})^2+1}.
\end{align}
\begin{align}\label{4.40f}
\sin^2{\psi}=\frac{r^2}{r^2+\dot{r}^2}=\left[1+\frac{1}{u^2}\left(\frac{du}{d\theta}\right)^2\right]^{-1}.
\end{align}
\begin{align}\label{4.40g}
\sin^2{\psi}=\left[1+\frac{e^2\sin^2{\theta}}{(1+e\cos{\theta})^2}\right]^{-1}.
\end{align}

\hspace*{5 mm} Now we look at the frequency term appearing in Eq.~\eqref{2.1e} which we call the de Broglie force. This relation was developed for massive particles but when we introduce light parameter $v=c$, it can be shown to reduce to the same formula that JPL used for determining anomalous acceleration of Pioneer 10. Here $d\nu/dt$ will be Doppler frequency drift $\dot{f}_p=f_0a_p/c$. Then de Broglie force can be shown to equal $hf_0a_p/c^2=ma_p$ where m is the equivalent relativistic mass of the DSN reference signal.

We have the fundamental relations $E=mc^2=h\nu$.
\begin{align}\label{2.2a}
\frac{m}{m_0}=\frac{\nu}{\nu_0}=\frac{dt}{d\tau}=\gamma=\frac{1}{\sqrt{1-\beta^2}}.
\end{align}
\begin{align}\label{2.3a}
\frac{1}{\nu_0}\frac{d\nu}{dt}=\frac{1}{m_0}\frac{dm}{dt}=\frac{d}{dt}\left(\frac{dt}{d\tau}\right)=\frac{d\gamma}{dt}\approx\frac{v}{c^2}\frac{dv}{dt}.
\end{align}
Therefore above relations show that the change in associated de Broglie frequency is same as the change in relativistic mass of the particle. Altered energy level of a body would imply altered relativistic mass of the body. Since in our theory, the gravitational attraction exists between the relativistic masses, the attractive force between the two bodies would also be altered. The third term in Eq.~\eqref{2.3a} would imply rate of change in the ratio of the coordinate time and the proper time. The fourth term would imply a rate of change in the flat Minkowski metric and the last term relates all above changes to the rate of change of the velocity.

\hspace*{5 mm} Single most major difference between PR and the general relativity is that the later allows only one time specific deviation to the flat Minkowski metric which we call weak field approximation, where as PR allows different deviations to the flat Minkowski metric to suit different two body problems. Therefore there is no such thing as a spacetime fabric or rubber in PR \cite{20}.

\hspace*{5 mm} Now how do we apply these concepts to the problem at hand. First we need to imagine how many millions and billions of fundamental particles have gone into building the Pioneer spacecraft. As the spacecraft passes through the gravitational field of the sun, each of those massive particles (which are in bound state) undergo frequency shift according to the scheme $d\nu/dt\neq0$ (de Broglie force $\neq0$) in our theory. The modern-day deep space navigational codes \cite{3} are obviously based on the notion $d\nu/dt=0$. This process of frequency shift of millions or billions of massive particles alters the energy level of the spacecraft. When the spacecraft is moving away from the Sun, its energy level goes down. When it moves towards the Sun, its energy level goes up. This causes a change in the relativistic mass of the spacecraft. This is reflected in the change of its kinetic energy. When moving away from the sun it losses its velocity compared to the model velocity and while moving towards the Sun it gains additional velocity. This is consistent with the direction of anomalous acceleration which is always sunwards. Since this process does not involve any change in the rest mass of the spacecraft and since we have introduced the concept of dynamic WEP (weak equivalence principle) which states that the gravitational mass is equal to the relativistic mass, there is no violation of the principle of equivalence. This aspect of the theory is discussed in little more detail elsewhere \cite {17}.\\
\hspace*{5 mm} When the spacecraft is near the Sun or a palnet in a stable near circular orbit, the frequency shift stops or is negligible. In this situation no anomalous acceleration would be detected. For the same reason no effect due to such anomalous acceleration could be detected on the orbital motion of the solar system planets \cite{8}. Therefore the anomalous acceleration could be observed only for those bodies that travel radially toward and away from the Sun like the hyperbolic trajectories of Pioneer 10/11 spacecrafts.\\
\hspace*{5 mm} The above explanation seems to fit well with the satellite flyby anomaly \cite{1} as well. It has been observed that satellites after an Earth swing–by possess a significant unexpected and unexplained velocity increase by a few mm/s. As the satellite comes closer to earth, it experiences stronger frequency shift of its constituent particles which significantly increases the energy level of the satellite resulting in increased velocity. The same explanation goes well for the observation concerning the return time of comets \cite{1}. Comets are observed to come back a few days earlier than predicted by the ordinary equations of motion. While on their outward journey from perihelion to aphelion, due to the frequency shift and consequent loss of energy, the comet would reach the aphelion earlier than expected because the aphelion would get shifted towards the Sun. Consequently the perimeter of the orbit would be substantially reduced. As it approaches the Sun it will gain the velocity as if nothing has happened. This reduced perimeter of the orbit could cause it to return earlier. This model however needs to be supported analytically. The second possible cause for the satellite flyby anomaly and the early return time of comets could be the factor  $\left(\cos{\psi}+\sin{\psi}\right)$ in Eq.~\eqref{2.1c}.

\hspace*{5 mm} As mentioned earlier, it is not very convinient to compute the gravitational frequency shift of millions and billions of massive particles but our theory proposes a definite connection between the proper time of the body and the gravitational frequency shift of its constituent particles, so the problem can be addressed by analyzing the proper time of the body as discussed in the following section. Same principles are used elsewhere to explain the rotation curves of the spiral galaxies \cite{32}. 

\section{Equations of spacecraft motion}
Following is the extention of the theory developed earlier \cite{17} where we introduced deviation to the flat Minkowski metric due to the gravitational field in the form,
\begin{align}\label{2.1}
\left(\frac{dt}{d\tau}\right)^2=\gamma^{2n}=(1-\beta^2)^{-n},
\end{align}
\begin{align}\label{3.2}
d\tau=dt\left(1-\frac{nv^2}{2c^2}\right),
\end{align}
where $t$ is the coordinate time, $\tau$ the proper time of the orbiting body and $n$ is a real number. The corresponding line element in polar coordinates is,
\begin{align}\label{2.2}
ds^2=c^2dt^2-ndr^2-nr^2d\theta^2-n(r^2\sin^2{\theta})d\phi^2.
\end{align}
We showed that the line element Eq.~\eqref{2.2} satisfies Einstein's field equations for any constant value of $n$ or for $n=0$. We derived the expression for the two body system in a gravitational field,
\begin{align}\label{3.4}
\frac{d^2\mathbf{r}}{d\tau^2}=
-\frac{\mu}{r^2}\left(1+(n-1)\frac{v^2}{c^2}\right)\mathbf{\hat r}.
\end{align}
In case of Pioneer spacecraft, we can identify the additional second order term in Eq.~\eqref{3.4} as the anomalous acceleration $a_p$ so that
\begin{align}\label{3.5}
-a_p\mathbf{\hat r}=-\frac{\mu}{r^2}(n-1)\frac{v^2}{c^2}\mathbf{\hat r}.
\end{align}
As pointed out earlier, unlike general relativity, PR allows introduction of different deviations to the flat Minkowski metric for different two body problems. Accordingly for spacecrafts subjected to anomalous constant sunward accelerations, such as Pioneer10/11, Galileo and Ulysses, we define the deviation factor $n$ as
\begin{align}\label{3.6}
n=\left(1+\frac{(\delta/2)}{(\mu/r^2)}\right),
\end{align}
where $\delta=|\delta \mathbf{\hat r}|$ is the unit acceleration. Substitution in Eq.~\eqref{3.5} gives
\begin{align}\label{3.6a}
-a_p=-\frac{\delta}{2}\frac{v^2}{c^2}.
\end{align}
The Pioneer 10 spacecraft had an approximate constant velocity of 12800 m/s at 40AU \cite{26}. This gives $-a_p=-9.115\times10^{-10}\: m/s^2$. The Pioneer 10 spacecraft had an approximate constant velocity of 12200 m/s at 67AU \cite{3}. This gives $-a_p=-8.2803\times10^{-10}\: m/s^2$. The Pioneer 11 spacecraft had an approximate constant velocity of 11600 m/s at 40AU \cite{3}. This gives $-a_p=-7.486\times10^{-10}\: m/s^2$. These values are well within the experimental accuracy of the observed values $-a_p=-(8.74\pm1.33)\times10^{-10}\: m/s^2$.

Substitution of Eq.~\eqref{3.6} in Eq.~\eqref{3.2} gives
\begin{align}\label{3.7}
d\tau=dt\left(1-\frac{v^2}{2c^2}-\frac{\delta v^2r^2}{4c^2 \mu}\right),
\end{align}
For the Pioneer 10 spacecraft velocities given above, the proper time intervals at 40AU and 67AU are 
\begin{align}\label{3.8}
d\tau=dt\left(1-9.115\times10^{-10}-1.2296\times10^{-4}\right),
\end{align}
\begin{align}\label{3.9}
d\tau=dt\left(1-8.28\times10^{-10}-3.134\times10^{-4}\right),
\end{align}
respectively. The last term is significant compared to the middle term. This amounts to 123 $\mu sec/sec$ lapse of time at 40AU and 313 $\mu sec/sec$ at 67AU. In comparision the GR formula gives
\begin{align}\label{3.10}
d\tau=dt\left(1-\frac{3\mu}{2c^2R}\right).
\end{align}
\begin{align}\label{3.11}
d\tau=dt\left(1-3.7\times10^{-10}\right),
\end{align}
\begin{align}\label{3.12}
d\tau=dt\left(1-2.21\times10^{-10}\right),
\end{align}
at 40AU and 67AU respectively. This is significantly different and would imply a drift in the proper time of the spacecraft. PR does not predict the drift but could allow such a drift provided it is experimentally verified. So the proper time parameter is a free parameter in PR. The lapse of time can be experimentally verified by measuring the drop in frequency of the atomic clock in the rest frame of the spacecraft or by measuring the drop in Doppler frequency in the rest frame of the spacecraft. In case of Pioneer 10 the drop in frequency of the atomic clock is expected to be $123\times10^{-6}\times9192631770=1130694$ Hz at 40AU and $2877294$ Hz at 67AU. For DSN signal of $2.11$ GHz, this drop is $259530$ Hz at 40AU and $660430$ Hz at 67AU.  

We can further generalize Eqs.~\eqref{3.6a} and ~\eqref{3.7} by introducing an approximate expression for the spacecraft velocity.
The vector constant called angular momentum per unit mass $\mathbf{h}$ is defined as
\begin{align}\label{3.12a}
\mathbf{h}=\frac{\mathbf{L}}{m}=\frac{\mathbf{p}\boldsymbol{\times}\mathbf{r}}{m}\equiv\frac{|\mathbf{p}||\mathbf{r}|\sin{\psi}}{m}\mathbf{\hat h}.
\end{align}
Using Eqs.~\eqref{3.12a} and ~\eqref{4.40g} we can write
\begin{align}\label{3.12b}
v^2=\frac{h^2}{r^2\sin^2{\psi}}=\frac{\mu^2}{h^2}(1+e\cos{\theta})^2\left[1+\frac{e^2\sin^2{\theta}}{(1+e\cos{\theta})^2}\right].
\end{align}
\begin{align}
v^2&=\frac{\mu^2}{h^2}(1+e^2+2e\cos{\theta})\label{3.12c}\\
&=\frac{\mu^2}{h^2}(2+2e\cos{\theta}+e^2-1)\label{3.12d}\\
v^2&=\mu\left(\frac{2}{r}+\frac{1}{a}\right)\label{3.12e}.
\end{align}
The above equation for hyperbolic trajectory is similar to the vis-viva equation for elliptical orbits of planets except that the quantities in brackets have positive sign between them. This expression for velocity may not be as precise as that obtained from  Doppler tracking or radar ranging techniques \cite{29} with N-body consideration but it seems to fit well with the Pioneer data. We can also derive the same expression using conservation of energy criteria where the semi-latus rectum to be used for hyperbolic trajectory is $a(e^2-1)$ and distance to periapsis is $a(e-1)$. We can introduce Eq.~\eqref{3.12e} in Eqs.~\eqref{3.6a} and ~\eqref{3.7} to get
\begin{align}\label{3.12f}
-a_p=-\frac{\delta \mu}{2c^2}\left(\frac{2}{r}+\frac{1}{a}\right),
\end{align}
\begin{align}\label{3.12g}
d\tau=dt\left[1-\frac{\mu}{2c^2}\left(\frac{2}{r}+\frac{1}{a}\right)\left(1+\frac{\delta r^2}{2 \mu}\right)\right].
\end{align}
In order to justify Eq.~\eqref{3.12f} one has to experimentally verify Eq.~\eqref{3.12g}.

Introduction of Eq.~\eqref{3.5} into Eq.~\eqref{3.4} gives
\begin{align}\label{3.13}
\frac{d^2\mathbf{r}}{d\tau^2}=-\frac{\mu}{r^2}\mathbf{\hat r}-a_p\mathbf{\hat r}.
\end{align}
Eq.~\eqref{3.13} can be written as  
\begin{align}\label{3.14}
\left(\frac{d^2 r}{d\tau^2}-\frac{h^2}{r^3}\right)\mathbf{\hat r}=
-\frac{\mu}{r^2}\mathbf{\hat r}-a_p\mathbf{\hat r}.
\end{align}
Substitution of following and Eq.~\eqref{3.6a} gives a second order non-homogeneous, non-linear differential equation.
\begin{align}\label{3.15}
u=\frac{1}{r}, \quad \text{and} \quad \frac{d}{d\tau}=hu^2\frac{d}{d\theta}. 
\end{align}
\begin{align}\label{3.15a}
\frac{d^2u}{d\theta^2}+u=\frac{\mu}{h^2}+\frac{a_p}{h^2u^2}.
\end{align} 
Here last term on right acts as a very small perturbation so when $a_p$ is set to zero, there remains no difference between the above equation and Newton's equation of motion. Further
\begin{align}\label{3.16}
\frac{d^2u}{d\theta^2}+u=\frac{\mu}{h^2}+\frac{\delta v^2}{2c^2h^2u^2}.
\end{align}
\begin{align}\label{3.17}
\frac{d^2u}{d\theta^2}+u=\frac{\mu}{h^2}+\frac{\delta}{2c^2\sin^2{\psi}}\left(\frac{v^2r^2\sin^2{\psi}}{h^2}\right).
\end{align}
It should be noted that $\theta$ in Eq.~\eqref{3.17} corresponds to $\phi$ in Eq.~\eqref{2.2}. As discussed in \cite{17} the quantity in brackets is unity. 
Introducing Eq.~\eqref{4.40f} and $u=(\mu/h^2)(1+e\cos{\theta})$ on r.h.s. of Eq.~\eqref{3.17} as a close approximation we get
\begin{align}\label{3.19}
\frac{d^2u}{d\theta^2}+u=\frac{\mu}{h^2}+\frac{\delta}{2c^2}+\frac{\delta e^2\sin^2{\theta}}{2c^2(1+e\cos{\theta})^2}.
\end{align}
\begin{align}\label{3.19a}
\begin{split}
\left[\frac{d^2u}{d\theta^2}+u\right](1+e\cos{\theta})^2=&\left[\frac{\mu}{h^2}+\frac{\delta}{2c^2}\right](1+e\cos{\theta})^2\\
&+\frac{\delta e^2}{2c^2}\sin^2{\theta}.
\end{split}
\end{align}
\begin{align}\label{3.19b}
\begin{split}
\frac{d^2u}{d\theta^2}+u=&\left[\frac{\mu}{h^2}+\frac{\delta}{2c^2}\right](1+e\cos{\theta})^2+\frac{\delta e^2}{2c^2}\sin^2{\theta}\\
&-\left[\frac{d^2u}{d\theta^2}+u\right](2e\cos{\theta}+e^2\cos^2{\theta}).
\end{split}
\end{align}
Again introducing $u=(\mu/h^2)(1+e\cos{\theta})$ in the last term on r.h.s. and simplifying we get
\begin{align}\label{3.19c}
\frac{d^2u}{d\theta^2}+u=\frac{\mu}{h^2}+\frac{\delta}{2c^2}(1+e^2)+\frac{\delta e}{c^2}\cos{\theta}.
\end{align}
The solution to Eq.~\eqref{3.19c} has the form
\begin{align}\label{3.19d}
u=C_1\cos{\theta}+C_2\sin{\theta}+\frac{\mu}{h^2}+\frac{\delta}{2c^2}(1+e^2) +\frac{\delta e}{2c^2}(\theta\sin{\theta}).
\end{align}
Solution $u_0$ of Eq.~\eqref{3.19c} when $a_p=0$ is given by the first three terms of Eq.~\eqref{3.19d}. Hence the correction due to $a_p$ is given by 
\begin{align}\label{3.19e}
(u-u_0)=\frac{\delta}{2c^2}(1+e^2+e\theta\sin{\theta}).
\end{align}
Eq.~\eqref{3.19e} can be exclusively used for deep space probe trajectories provided the corresponding proper time Eq.~\eqref{3.7} is experimentally verified. In case of the Pioneer 10 spacecraft, $(u-u_0)$ is of the order $3.97\times10^{-17}$. 

The theory given above is also approximately supported by Einstein's field equation as discussed earlier \cite{17}.  We replace factor $(n-1)=\xi$ in Eq.~\eqref{3.4} by a constant $\xi=150000$ for Pioneer 10 spacecraft. This will give us the value of $n=150001$. This constant value of $n$ will satisfy Einstein's field equations as per the first of the conditions discussed in \cite{17} and given in Eq.~\eqref{3.20}. 
\begin{align}\label{3.20}
\left(\frac{r}{n}\frac{\partial n}{\partial r}\right)=0 \qquad and \qquad \left(\frac{r}{n}\frac{\partial n}{\partial r}\right)=-4.
\end{align}
Using Eq.~\eqref{3.5} we get
\begin{align}\label{3.21}
-a_p\mathbf{\hat r}=-\frac{150000\mu v^2}{c^2r^2}\mathbf{\hat r}.
\end{align}
On 1st January 1987, the Pioneer 10 spacecraft was approximately 40AU from the Sun, and receding with a nearly constant velocity of 12800 m/s \cite{26}. So we get $a_p=1.0135\times10^{-9}$ m/s @ 40AU and $a_p=3.28\times10^{-10}$ m/s @ 67AU for $v=12200 m/s$.
For Pioneer 11 spacecraft the applicable constant is $\xi=80000$. In November 1995, the Pioneer 11 spacecraft was approximately 40AU from the Sun receding with a nearly constant velocity of 11600 m/s. So we get $a_p=4.43\times10^{-10}$ m/s @ 40AU and $a_p=1.11\times10^{-9}$ m/s @ 27AU assuming $v=12400 \:m/s$. Hence we conclude that Einstein's field equations do provide explanation for Pioneer anomaly but only if the weak field approximation and the Schwarzschild solution are replaced by the present theory. Secondly, requirement that $\xi$ has to be a constant introduces unacceptable variation over the distances. For this reason it is proposed to go beyond Einstein's field equations and have a variable $n$ as defined in Eq.~\eqref{3.6}.

The second condition discussed in \cite{17} and given in Eq.~\eqref{3.20} that would allow the Ricci tensor to vanish and satsfy Einstein's field equation gives variable $n$, but it also gives unacceptable variation over distances. If we integrate second condition in Eq.~\eqref{3.20} we get
\begin{align}\label{3.22}
\int \frac{\partial n}{n}=-4 \int \frac{\partial r}{r}.
\end{align}
\begin{align}\label{3.23}
\ln{n}-\ln{C}=-4 \ln{r}.
\end{align}
\begin{align}\label{3.24}
(n-1)=\left(\frac{C}{r^4}-1\right).
\end{align}
So Eq.~\eqref{3.21} gets replaced by
\begin{align}\label{3.25}
-a_p\mathbf{\hat r}=-\frac{\mu v^2}{c^2r^2}\left(\frac{C}{r^4}-1\right)\mathbf{\hat r}.
\end{align}
Here $C$ is a constant of integration. No matter how we define this constant, Eq.~\eqref{3.25} yields large variations for $a_p$ over distances. For example, for an initial condition of $a_p=8.74\times10^{-10}\:m/s^2$ at 40AU with constant $v=12800\:m/s$ we get $C=1.65858377908\times10^{56}$. Using this constant for 67Au and $v=12200 \:m/s$ we get $a_p=3.595\times10^{-11}\:m/s^2$. This shows that Einstein's field equations in association with the line element Eq.~\eqref{2.2} does provide some clue to the Pioneer anomaly but the solutions involve unacceptable variation over distances. The solution to Pioneer anomaly given by Eq.~\eqref{3.6a} gives $a_p$ within the limits of observational accuracy. We can compare this solution with the conditions ~\eqref{3.20} imposed by Einstein's field equations. If we differentiate Eq.~\eqref{3.6}, we can write
\begin{align}\label{3.26}
\left(\frac{r}{n}\frac{\partial n}{\partial r}\right)\approx2-4\frac{\mu}{\delta r^2}=2\left(\frac{n-2}{n-1}\right).
\end{align}

\section{Conclusion}
PR proposes alternate ways of deviating from the flat Minkowski metric which are related to physical phenomenon. This is possible because such unconventional deviations does not affect the fundamental derivations of gravitational redshift of light, bending of light and perihelic precession of planets. This is not possible in general relativity. Many of the theories that remain within the framework of general relativity have to deviate from fundamental physics elsewhere in order to provide solution for the anomaly.\\
\hspace*{5 mm} The theory simultaneously explains the anomalous accelerations of Pioneer 10/11, Galileo and Ulysses spacecrafts. The theory points out at the limitations of the weak field approximation and proposes a drift in the proper time of the spacecraft outside the framework of general relativity. Usually general relativity theories speak of drift in the proper time in relation to the expansion of univrse or quadratic drift of the earth clocks or accelerated Sun etc. In PR the proper time of a body is associated with the gravitational frequency shift of the constituent fundamental particles of the body. The frequency shift changes the energy level of the body which gets reflected in its relativistic mass and therefore in its motion. This change in energy level causes the time like geodesics to deviate from that of the standard theoretical models. We introduce proper time in the line element of a metric theory according to a fixed set of rules laid down by general relativity for introducing deviation in the flat Minkowski metric. The frequency shift for bodies of different composition traversing different trajectories however, is not the same and this gets reflected in its motion as an unmodeled anomalous effect. This association of proper time with the gravitational frequency shift of the body requires the flat Minkowski metric to deviate in different ways for different two body systems. This solves the problem of anomalous acceleration in a very simple way and yields equations of motion that account for the anomalous acceleration. Analytical determination of the change in energy level of the spacecraft due to gravitational frequency shift of its constituent fundamental particles can be the next step towards further verification of this theory. It is likely that the Pioneer anomaly provides the proof of existence of the de Broglie force. The solution to Pioneer anomaly given by Eq.~\eqref{3.6a} gives $a_p$ within the limits of observational accuracy. Introduction of dynamic WEP in PR assures that the principle of equivalence is not violated in case Pioneer anomaly.

\section{Acknowledgement}
Author is grateful to John D. Anderson for providing Pioneer 10 trajectory data and for helpful discussion about the angle between the radial and velocity vectors and about the spacecraft velocities. Author is also thankful to John Hodge for comments and suggestions.
\pdfbookmark{References}{Ref}
\bibliographystyle{amsalpha}

\end{document}